\documentclass[trackchanges,twocolumn]{aastex7}
\usepackage{graphicx}
\usepackage{gensymb}
\usepackage{longtable}
\usepackage{amsmath}

\def\O3{[\ion{O}{3}]}

\begin{document}


\title{The Ejecta Nebula Around the Wolf-Rayet Star WR 71}

\author[0000-0003-3829-2056]{Robert A.\ Fesen}
\email{robert.fesen@dartmouth.edu}
\affil{6127 Wilder Lab, Department of Physics and Astronomy, Dartmouth
       College, Hanover, NH 03755 USA}

\author{Manuel C.\ Peitsch}
\email{manuelcp.astro@gmail.com}
\affiliation{Roboscopes at e-EyE Entre Encinas y Estrellas Observatory, 
Camino de los Molinos 06340 Fregenal de la Sierra, Spain}

\author{Martin R.\ Pugh}
\email{mpastro2001@yahoo.co.uk}
\affil{Observatorio El Sauce, 
Rio Hurtado, Coquimbo, Chile}

\author{Bray Falls}
\email{brayfalls@gmail.com}
\affiliation{Sierra Remote Observatories, 42120 Bald Mountain Road, Auberry, CA, 93602, USA}
\affiliation{Starfront Observatories, 1724 Co Rd 244, Rockwood, TX 76873, USA}

\author[0000-0002-7855-3292]{Marcel Drechsler}
\email{epost@marcel-drechsler.de}
\affiliation{\'Equipe StDr, B{\"a}renstein, Feldstraße 17, 09471 B{\"a}renstein, Germany}

\begin{abstract}

\noindent
We present deep H$\alpha$ and [\ion{O}{3}] images
of the ejecta rich nebulosity associated with the suspected runaway and binary Wolf-Rayet star WR~71 (HD 143414).
In H$\alpha$ emission, the nebula appears as a crescent shaped,
broken ring of clumpy emission some $9' \times 13'$
in angular size centered to the south and west of the WR star. At a Gaia estimated distance of 4.27 kpc, the nebula has physical dimensions of $11 \times 16$ pc making it one of the larger known ejecta rich WR ring nebulae.
Our \O3 image also show considerable surrounding faint nebulosity much of which may be unrelated to the WR star. A comparison of the
nebula's optical appearance with that seen in WISE 22 $\mu$m data shows infrared coincidence with the nebula's brightest \O3 emission features. Deep H$\alpha$ and \O3 images like those presented here suggests that new and substantially deeper imaging reconnaissance of WR star nebulae compared to earlier surveys may lead to additional WR ring nebula detections, thereby enhancing our understanding on the frequency and formation of WR ring nebulae.

\end{abstract}


\bigskip


\section{Introduction}

Emission line sky surveys have lead to the 
discovery of hundreds of extended emission line objects including H~II regions, planetary nebulae, 
and supernova remnants. One of the rarer class of emission line objects is that associated with Wolf-Rayet (WR) stars.
These stars are among a galaxy's
most massive stars with initial masses $\geq30~$M$_{\odot}$ 
\citep{Maeder1994, Conti2000, Crowther2008}.
With stellar wind velocities ranging 
from 1000 to 4000 km s$^{-1}$ plus very high luminosities, these stars undergo high  mass-loss rates $\sim$ $10^{-5}$ M$_{\odot}$ yr$^{-1}$ 
\citep{Hamann2019,Sander2022}.

Divided into WN and WC types based 
on their spectra showing either 
broad emission lines of helium and nitrogen ions (WN), or carbon and oxygen lines in addition to helium lines (WC), they are further subdivided  into excitation/ionization
sequences where the highest to lowest excitation spectral features are denoted by low to high subclass numbers.
WN stars are arranged WN3 to WN9 based on the
relative strengths of \ion{N}{3}, \ion{N}{4}, and
\ion{N}{5} lines, whereas  WC stars are subtyped 
based on excitation/ionization differences, ranging from WC3 to WC9. 

Due to their strong stellar winds and high mass loss rates, 
WR stars are often seen surrounded by optical emission arcs, rings or  shells \citep{Chu1981, Chu1983, Miller1993}.
Such WR star associated optical nebulae are categorized based on their formation mechanisms: 
W-type for a wind-blown ISM bubble, E-type for stellar mass ejection nebulae, and  R-type for a radiatively excited H~II region.
Most WR nebula are associated with WN type stars, and WN8 stars in particular 
\citep{Chu2016}. 

The relatively rare E-type WR star ring nebulae are especially interesting due to their
formation by means of substantial and violent mass loss episodes in the recent history of the star's mass loss evolution. Unlike wind-blown or H~II region types, ejecta WR nebulae tend to
more clumped and irregular in morphology
\citep{Chu1981, Chu1991}. Only a few are known and these show higher nitrogen and helium abundances relative to the ISM consistent with CNO cycle processed material \citep{Kwitter1984, Esteban2016}.

\begin{deluxetable*}{lcccccccl}
\tablecolumns{9}
\tablecaption{Comparison of Ejecta Type Wolf-Rayet Nebulae \label{Tab2}}
\tablewidth{0pt}
\tablehead{\colhead{WR } & \colhead{Nebula} & \colhead{Wolf-Rayet} & \colhead{Spectral} & \colhead{Nebula Dimensions} 
& \colhead{Distance\tablenotemark{a}} & \colhead{Physical Size} & \colhead{$|z|$}   & \colhead{References} \\
\colhead{Name}  & \colhead{Name}  & \colhead{Star}   & \colhead{Type}  & \colhead{(arcmin)}  & \colhead{(kpc)} & \colhead{(pc)} & \colhead{(pc)}     & \colhead{} }
\startdata
  WR 6   & S308       & HD 50896    & WN4      &  40  & $1.51\pm0.09$   &  17.6  &  265      & 1, 2, 3 \\
  WR 8   &   \nodata  &  HD 62910   & WN6/WC4  &  6   & $3.52\pm0.16$   &  6.1   &   230     & 1, 4  \\ 
  WR 40  & RCW 58     &  HD 96548   & WN8      &  7 x 9 & $2.70\pm0.12 $  & 5.5 x 7.1 & 230  &  5 \\ 
  WR 71 &  \nodata    & HD 143414   & WN6 + O?  &  9 $\times$ 13   & $4.27\pm 0.39$    & 11 $\times$ 16         & 565  & 1, 6, 7 \\
  WR 124 & M1-67      & BAC 209     & WN8      &  1.2 & $5.36\pm0.38$     &  1.9      &  310  &  5      \\
  WR 136 & NGC 6888   & HD 192163   & WN6      &  12 x 18 & $1.67\pm0.04$  &  5.8 x 8.7  & 70   & 1, 5, 8, 9 \\
\enddata
\tablenotetext{a}{WR star distances are Bayesian corrected geometric values from \citet{Bailer2021} for Gaia DR3. }
\tablenotetext{}{References: 1) \citet{Stock2010} ; 2) \citet{Esteban1992a} ; 3) \cite{Chu2003};
4) \citet{Fesen2025};  5) \citet{Chu1983}; 6) \citet{Marston1994b}; 7) this paper, 8) \citet{Chu1991} 
9) \citet{Esteban1992b}  }
\label{table}
\end{deluxetable*}

Table 1 compares properties of some  ejecta rich Galactic WR nebulae as complied by \citet{Chu1991}, \citet{Marston1997} and
\citet{Stock2010}. One of the least studied WR nebula is that associated with WR~71 which is a 10.1 V mag WN6 star also known as HD~143414. A faint H$\alpha$ nebula around WR~71 was first detected by \citep{Marston1994b} using the CTIO 0.6 m Curtis-Schmidt telescope. A better detection was later made by
by \citet{Stock2010} (Fig.\ 1) through an examination of 1.2 m Schmidt AAS/UKST SuperCOSMOS 
H$\alpha$ survey images of the southern galactic plane
also known as the Southern H$\alpha$ Survey or SHS
\citep{Parker2005}. Due to the nebula's highly clumpy appearance, both \citet{Marston1997} and \citet{Stock2010} classified the WR~71 nebula as an E or ejecta type WR ring nebula.

\begin{figure}[]
\centerline{\includegraphics[angle=0,width=8.0cm]{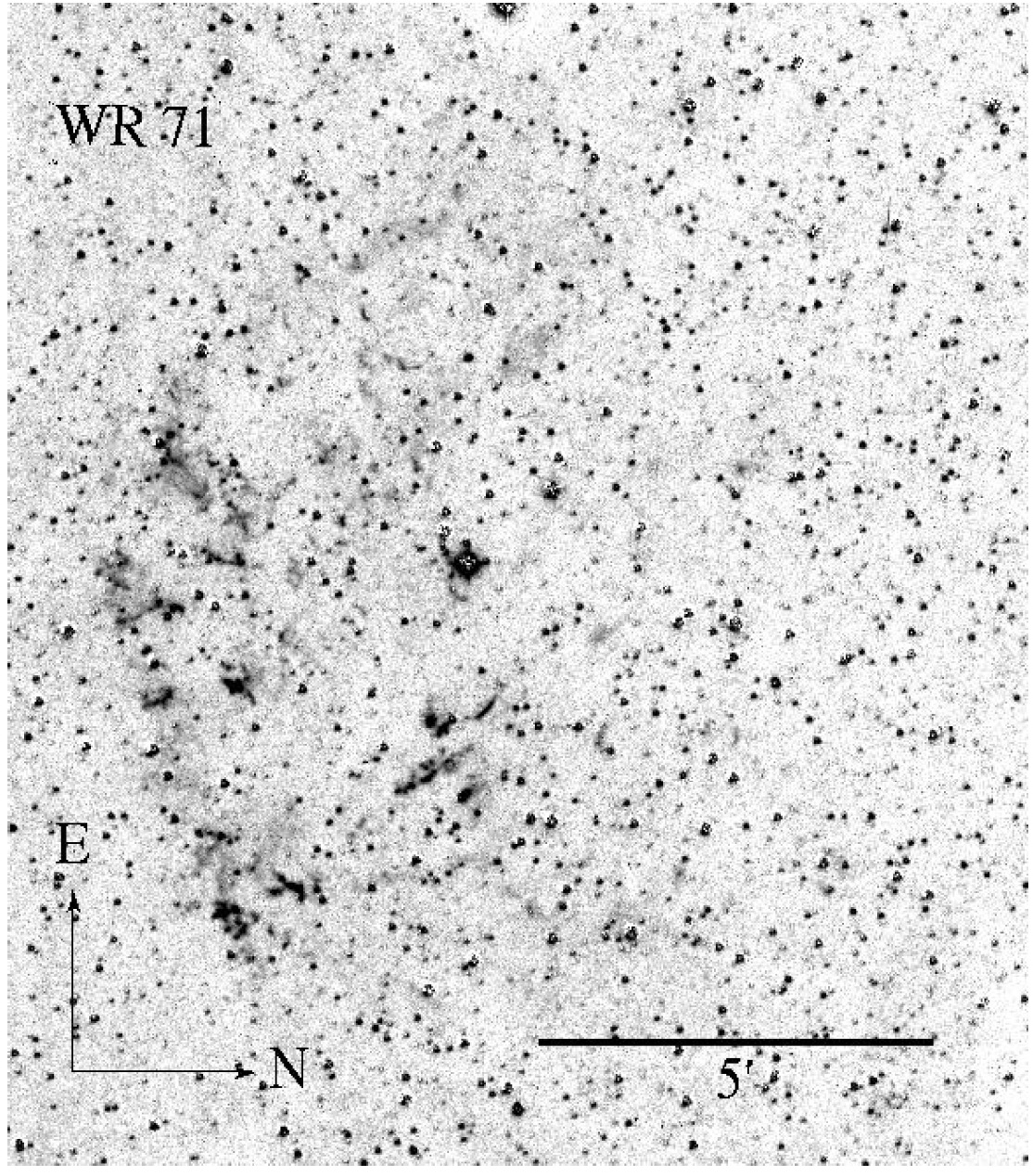}}
\caption{1.2 m Schmidt H$\alpha$ image showing nebulosity around WR~71 (from \citealt{Stock2010}).
\label{Stock}  }
\end{figure}

No follow-up imaging of the WR~71 nebula have been published. However, much deeper imaging of this WR star's nebulae is possible using even small telescopes with current detectors and modern digital imaging processing.
Because surface brightness is independent of telescope aperture, and because detected surface brightness is a function of pixel size on
the sky, the number of pixels in an object, and 
the exposure time, long series of exposures taken with small telescopes with large FOVs and pixel scales $\sim2''$ has been shown to detect faint emission from large, extended objects as good or even better than using large telescopes (e.g., 
\citealt{Java2016,Martinez2020}).

Moreover, recent developments in affordable large-format CMOS detectors plus 
high transmission (T $\geq95$\%) narrow passband 
filters (FWHM $\approx$ 30 \AA) has led to a revolution in deep emission-line imaging
of Galactic nebulae by amateurs. Not restricted by telescope allocation committees or outside funding, amateur astronomers taking literally hundreds of exposures with small telescopes equipped
with high-throughput emission-line filters and sensitive
digital detectors are capable of detecting previously unknown and extremely faint  Galactic emission line nebulae. 
Moreover, unlike previous emission sky surveys, amateurs have employed a variety of nebular emission line filters besides H$\alpha$, most importantly \O3 $\lambda$5007, and have imaged regions both in and far from the Galactic plane.

Here we present deep optical emission line images of the ejecta nebulosity associated with WR~71 obtained with amateur class telescopes and instruments.
Our observations are described in $\S2$ with our images presented and discussed $\S3$. 

\section{Observations}

Narrow passband H$\alpha$ and \O3 $\lambda$5007 filter images of WR~71 were obtained 
by M.R.P.\ in March and April 2025 using 
a 0.60 m PlaneWave CDK24 telescope at the Observatorio El Sauce located in Rio Hurtardo Valley, Chile.  These data were taken using 3 nm passband H$\alpha$ and [\ion{O}{3}] $\lambda$5007 filters located behind a 0.66 image reducer yielding an
f/4.37 focal ratio,  and a Moravian C1-61000 CMOS camera  binned 2x2 yielding
$4800 \times 3194$ pixels. This system provided an image scale of
$0\farcs59$ pix$^{-1}$ and a FOV of $47' \times 31'$.
Total exposure time in H$\alpha$ was 9.6 hr ($58 \times 600$ s) and 8.3 hr ($50 \times 600$ s) in [\ion{O}{3}].
Broadband
R,G,B filter images were also obtained with 42 minute 
exposures for each filter. These data were reduced using
commercial imaging processing software including Photoshop,  PixInsight and 
NoiseXTerminator. 

Follow-up images for line emission flux estimates were subsequently obtained in late April 2025 by M.C.P.\ using
a 0.43 PlaneWave CDK17 telescope also located at the Observatorio El Sauce. 
With an f/6.8 and an SBIG XSTL-11002 CCD camera with AO,
this system provided an image scale of $0.63''$ and a FOV of 
$41' \times 28'$.
Both  H$\alpha$ and \O3 images were obtained of WR~71 along with images of three flux calibration stars: PG~1323-086,
PG~1545+035, and G60-54 (Wolf~457) \citep{Massey1988, Oke1990,Stone1996}.

\begin{figure}[]
\centerline{\includegraphics[angle=0,width=8.8cm]{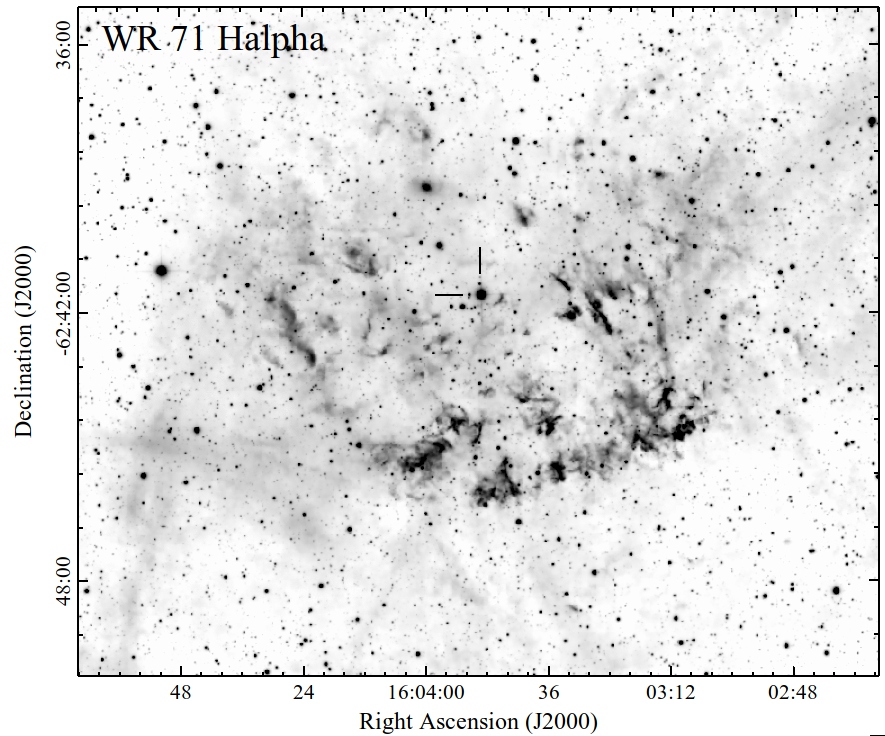}}
\centerline{\includegraphics[angle=0,width=8.8cm]{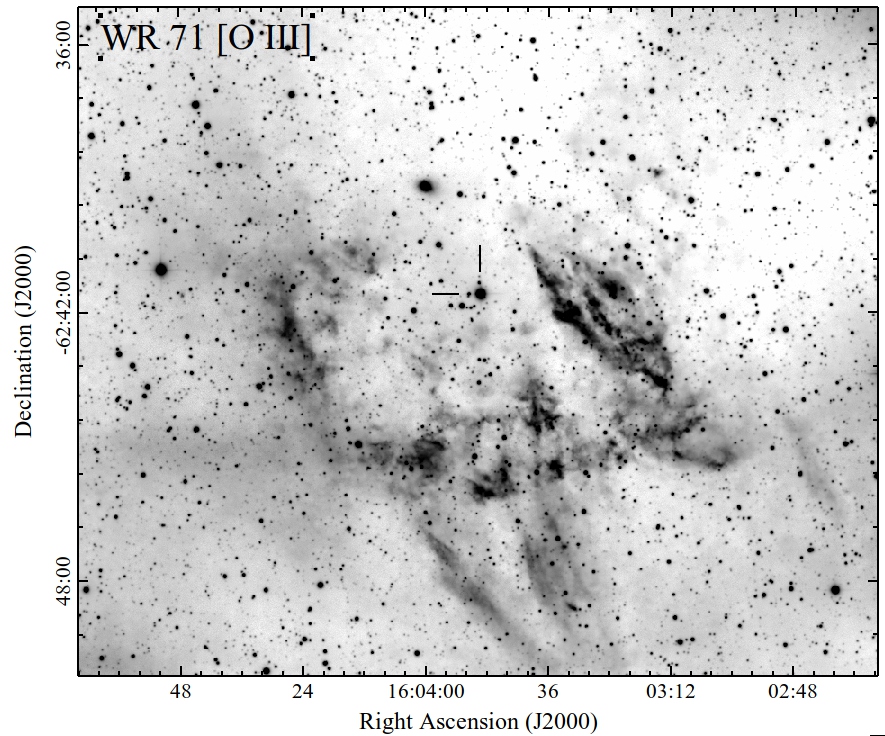}} 
\caption{Our deep H$\alpha$ and \O3 images of the WR~71 nebula. The Wolf-Rayet star, HD 143414, is marked.}
\end{figure}


\section{Results and Discussion}

Figure 2 shows our 
H$\alpha$ and [\ion{O}{3}] $\lambda$5007 images of WR 71. These reveal a more extensive nebulosity around and near the WR stars than seen in the H$\alpha$ Schmidt SHS image shown in Fig. 1. 
Broadly, the nebula appears as a 
broken ring of emission clumps not exactly centered on the WR star, HD 143414. In fact, the  bulk of the  nebula's H$\alpha$ emission lies to the south and west of the WR star in the direction of the star's Gaia derived proper motion 
(RA: $-6.4$ mas, Dec: $-11.6$ mas) in a crescent shape some 9$' \times 13'$ high and wide.   At a Gaia estimated distance of 
$4.27 \pm 0.38$ kpc, the nebula has physical dimensions of $11 \times 16$ pc making it one of the larger WR ring nebulae (see Table 1). Our images do not show evidence of a bow-shock nebula like that proposed for WR~71 by \citet{Meyer2020} but do reveal
considerable faint diffuse H$\alpha$ and \O3 emission in and around the nebula of which only some appears associated with the WR nebula.

\begin{figure*}[ht]
\begin{center}
\includegraphics[angle=0,width=8.1cm]{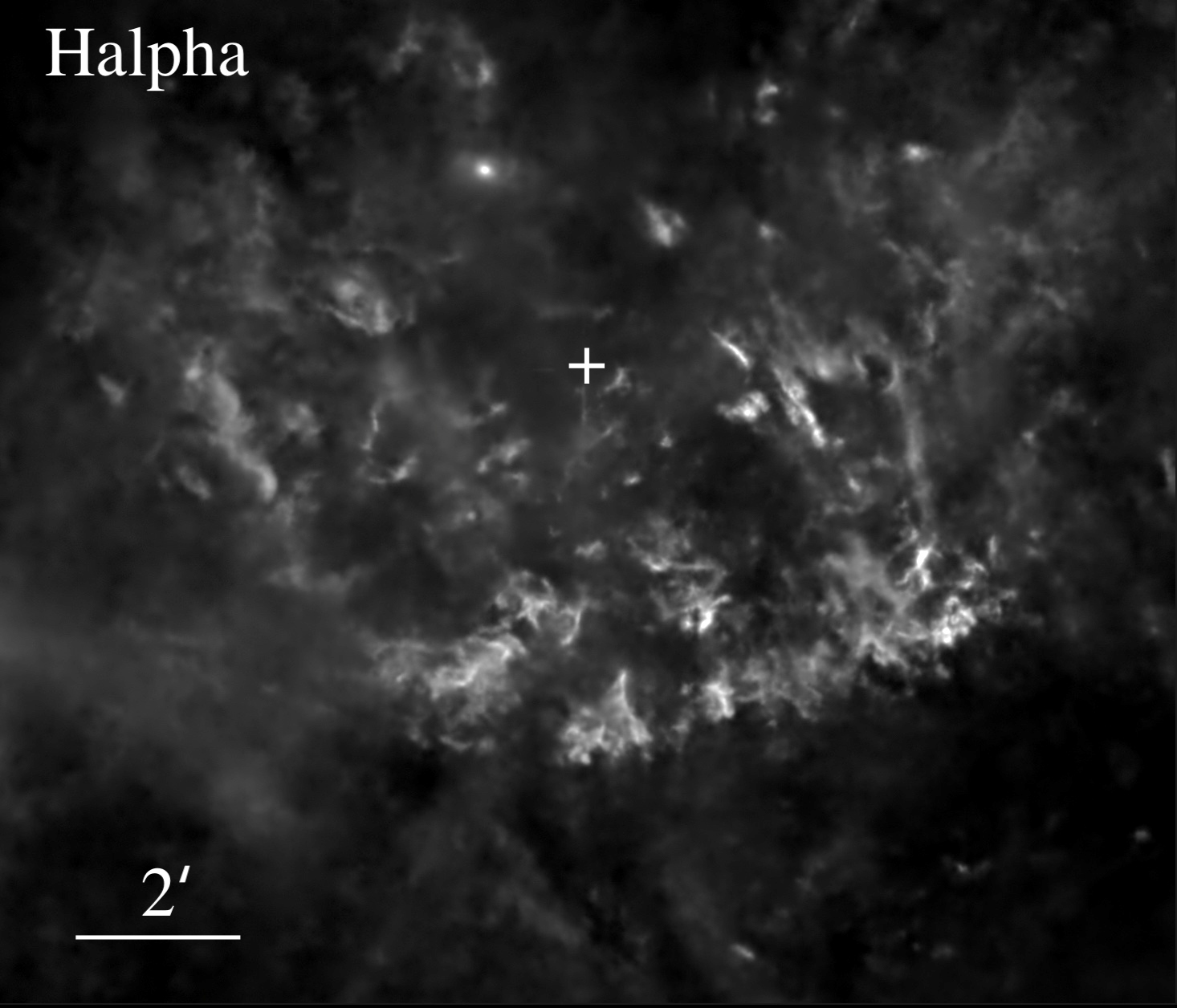}
\includegraphics[angle=0,width=8.05cm]{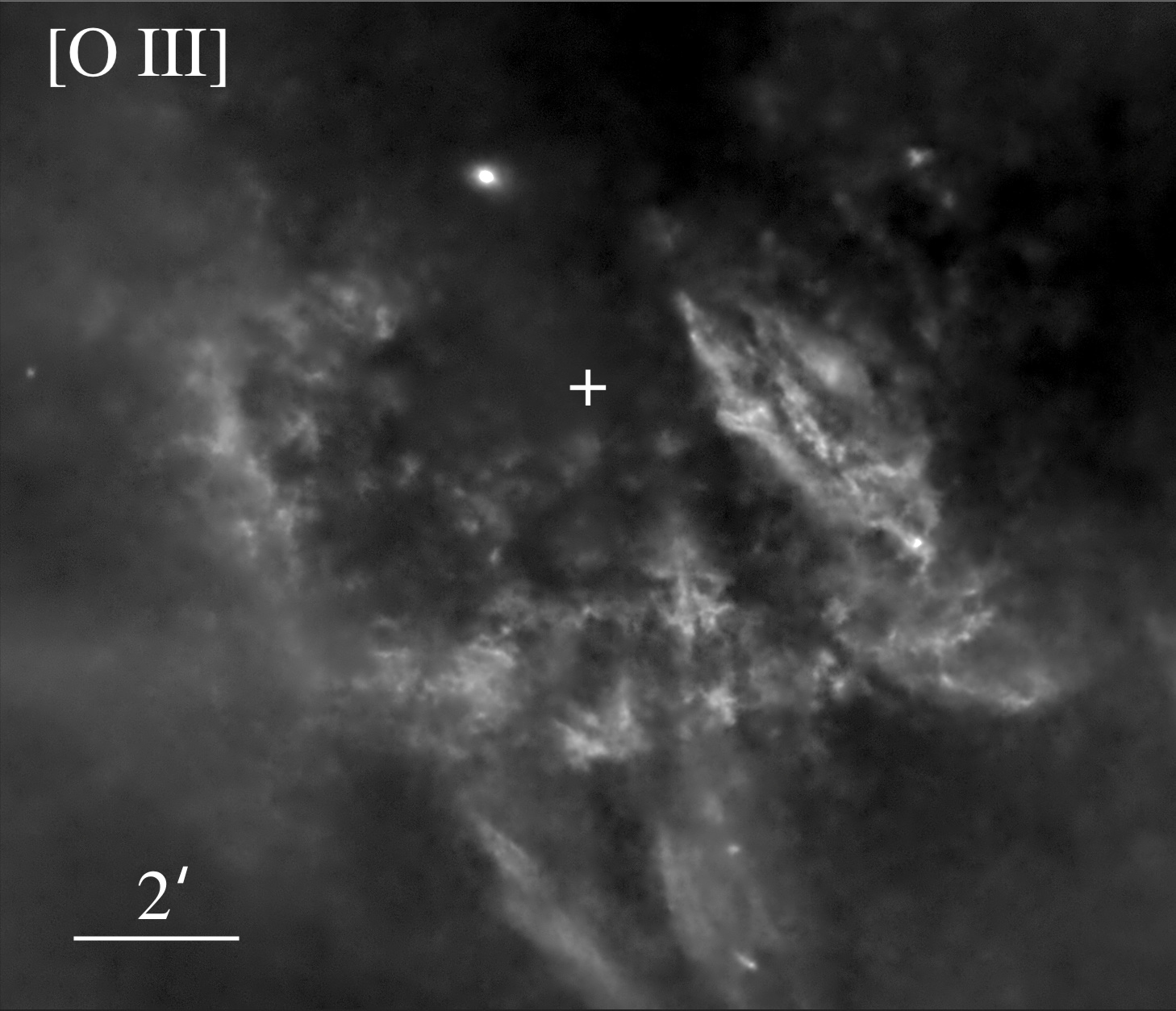}
\caption{Starless H$\alpha$ and [O III] images of the WR~71 nebula showing a clumpy
H$\alpha$ emission structure but one more diffuse in [O~III]. The crosses mark the WR star's location. The bright fuzzy object NE of the WR star is an elliptical galaxy.
\label{Fig4}
} 
\end{center}
\end{figure*}

The nebula's \O3 emission morphology is quite different from that seen in H$\alpha$.
The most notable differences are a lack of emission as far north of the WR star as seen in H$\alpha$ and the two broad emission regions extending some $4'$ southward from the nebula's bright southern shell.
Although it is tempting to connect these two broad \O3 emission features to WR~71, it is not certain that they are actually related to WR~71 as we will discuss below.
The nebula also appears 
a bit brighter in \O3 than in H$\alpha$ with strikingly bright east and western limbs compared to that seen in H$\alpha$.

The WR~71 nebula is classified as an ejecta type and the nebula's clumpy nature can be clearly seen in the enlarged starless images shown in Figure 3. While bright H$\alpha$ clumps are found along the nebula's southern and western regions, even the fainter emission features seen elsewhere exhibiting a clumpy structure. In contrast, a much more diffuse and patchy morphology is seen in the nebula's \O3  emission (Fig.\ 3, right panel.) With the exception of the bright western line of emission, the nebula's \O3 emission is diffuse in nature. The WR 71 star, HD 143414, lies north of the nebula's center in both H$\alpha$ and \O3 images, but this offset is much more noticeable in the \O3 image. 

Due to its $-150$ km s$^{-1}$  heliocentric radial velocity and early estimates $\sim 750 -1200$ pc  for its  distance off the Galactic plane  
\citep{Isserstedt1983, Stock2010, Moffat1998}, the WR~71 star, HD `143414,  has been viewed as a likely runaway star. 
However, at Galactic coordinates of $l = 323.1\degr, b = -7.6\degr$, its Gaia DR3 estimated distance of $4.27$ kpc yields a smaller z distance of around 560 pc (see Table 1).
Photometry and spectra indicate orbital period of 7.69 d with an unseen companion  
\citep{Isserstedt1983} but subsequent photometry by \citet{Balona1989} could not confirm this period, and
\citet{Hamann2006} have questioned the evidence for its binarity.

As seen in Figure 2, 
there is considerable faint diffuse emission around the WR 71 nebula. This is better seen in the wider FOV view of Figure 4 which presents a color composite image of the WR 71 nebula and its immediate surroundings. Our images show considerable H$\alpha$ and \O3 emission outside of the WR 71 nebula, extending far to the northwest and southeast.  The nature and distance of this emission is unknown, but the coincidence of the southern rim of the WR~71 nebula with the southern edge of this emission suggests a possible relation.

Furthermore,  faint filaments of \O3 emission
seen to the west from the WR~71 nebula raises questions about the connection of the bright \O3 emission filaments to the WR~71 nebula.
Indeed, the whole region toward WR~71 shows a variety of emission features.
The bright star to the northwest, a 6.3 mag B9.5 V star HD~143238 lying at a Gaia DR3 distance of 105 pc, 
exhibits an apparent reflection nebula in the form of a surrounding diffuse halo with a $\sim$20$'$ long tail extending to the southwest.

\begin{figure*}[h]
\begin{center}
\includegraphics[angle=0,width=15.0cm]{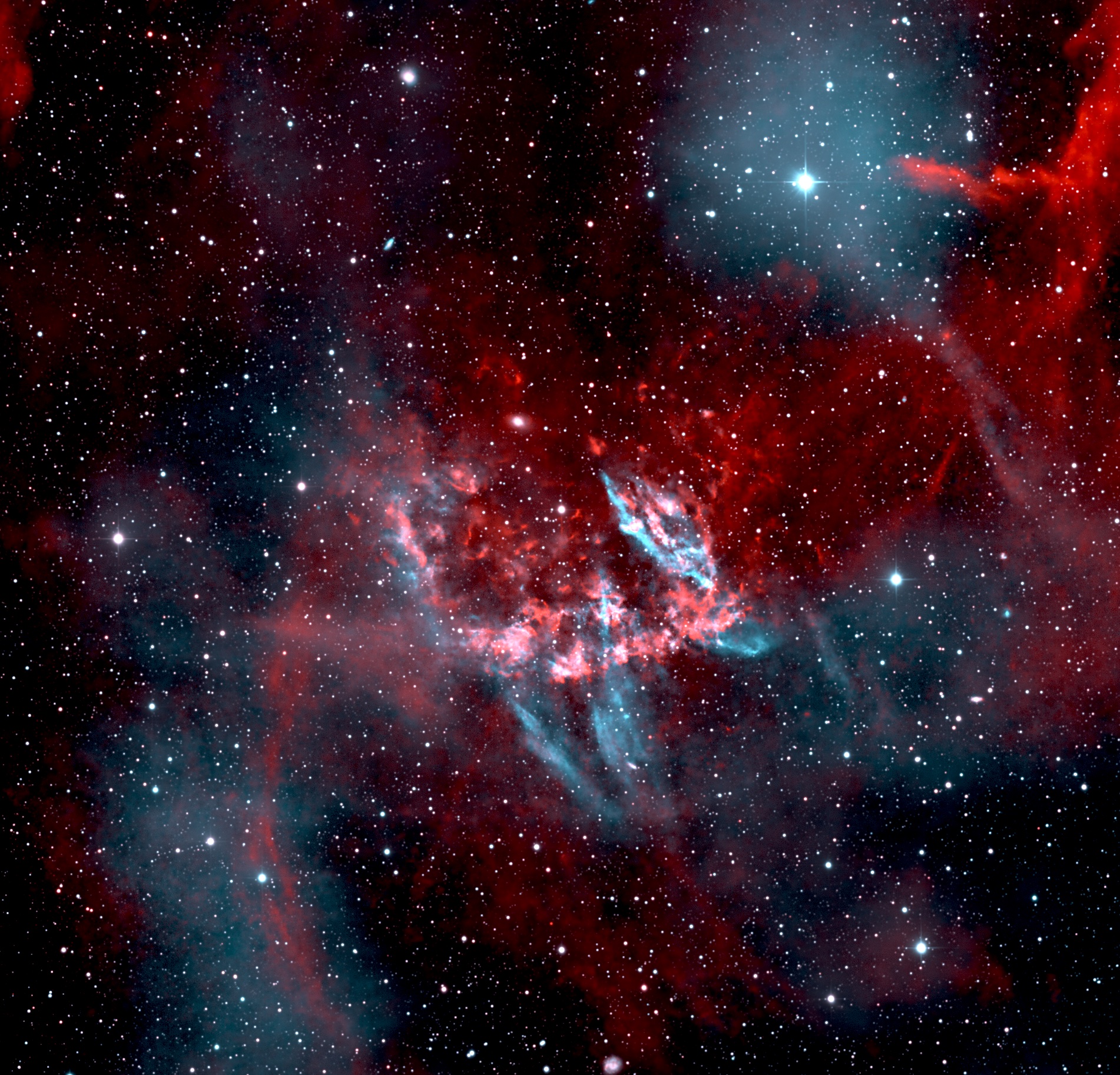}
\caption{Wide FOV color image of WR~71 composed of H$\alpha$ (red), 
[O~III] (blue) and broadband filter images.
\label{Fig4}
} 
\end{center}
\end{figure*}

\begin{figure*}[h!]
\begin{center}
\includegraphics[angle=0,width=15.0cm]{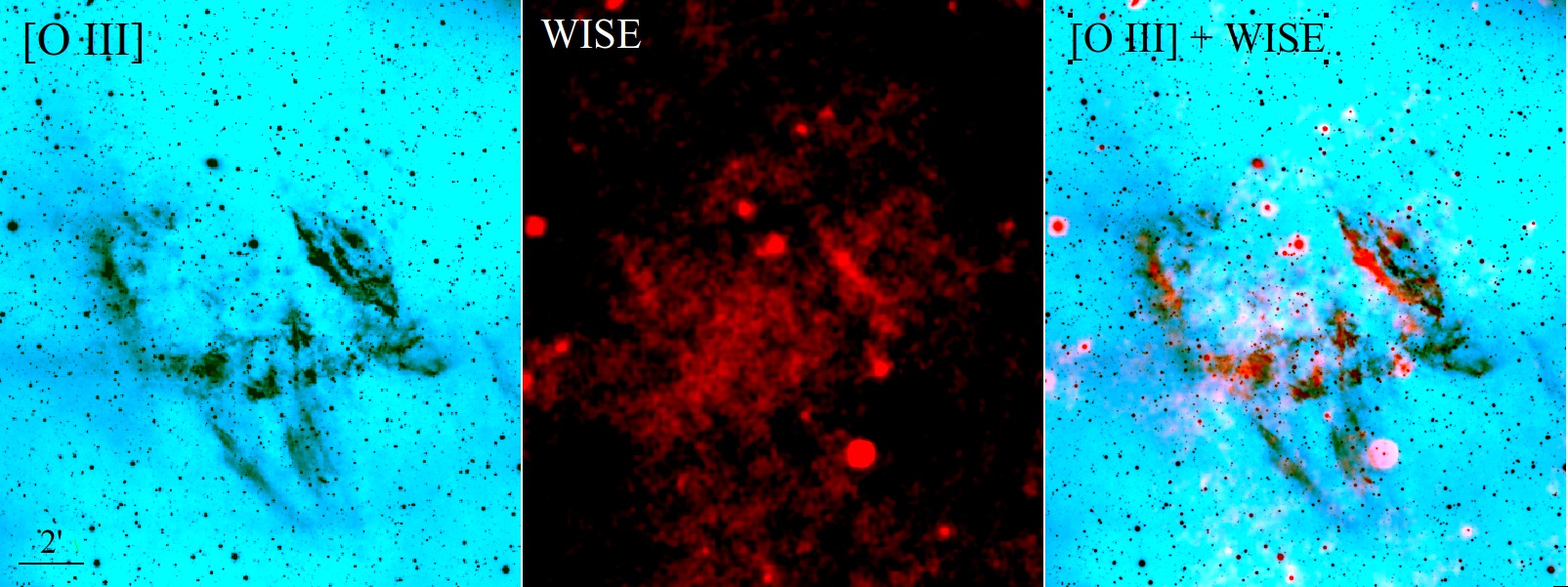}
\caption{Left Panel: Our [O III] $\lambda$5007 image shown in blue and green. Middle Panel: WISE 22 $\mu$m shown as red. 
Right Panel: Color composite of 
our [\ion{O}{3}] and the WISE 22
$\mu$m images. North is up, east to the left.
\label{Fig2}
} 
\end{center}
\end{figure*}

We note that WR~71's nebulosity is relatively faint, consistent with the lack of previous follow-up imaging. 
Our H$\alpha$ images are deeper than that the  $\leq$5 Rayleighs ($\leq$ $2.8 \times 10^{-17}$ erg cm$^{-2}$ s$^{-1}$ arcsec$^{-2}$) quoted as the detection limit of the SHS H$\alpha$ survey
\citep{Parker2005} which \citet{Stock2010} used to image
the WR~71 nebula (Fig.\ 1). 
Looked at another way, based on our calibration star observations and assuming 
an 45 \AA \ effective equivalent width for 
our H$\alpha$ filter and a nebula H$\alpha$ line width of $\simeq$0.8 \AA \ ($\simeq$ 40 km s$^{-1}$), we estimate an 
H$\alpha$ magnitude range for  
WR~71's emission of 21.8 mag to 23.5 mag arcsec$^{-2}$. 

WR 71's relative faintness may have be part of the reason why it was not included in the morphological study by \citet{Toala2015} of  WR stars in the infrared using 
Wide-field Infrared Survey Explorer (WISE) infrared images. Nonetheless, the WR 71 nebula does exhibit some 22 $\mu$m emission that positionally correlates qith the nebula's \O3 emission. This is shown in Figure 5. 
While there is considerable diffuse 22 $\mu$m emission
toward WR~71, the two brightest infrared WISE 22 $\mu$m filaments align well with
WR~71's nebula's bright east and west \O3 filaments
(Fig.\ 5, right panel). \citet{Toala2015} cited evidence from {\sl Spitzer} spectra that indicated 
emission in the WISE 22$\mu$m band is dominated by thermal emission from dust suggesting that the hotter regions
of the WR~71 nebula are those regions brightest in \O3 emission.

In summary, our images
reveal a much more extensive WR~71 emission nebula than previously 
realized. They also show the presence of 
additional faint diffuse emission in the direction to the WR star whose nature is presently unclear. However, now
realizing the presence of this additional emission, future spectroscopy could investigate this emission to see if it is somehow related to WR~71 nebula or the WR star's earlier mass loss.

The scientific value of investigating WR star ring nebulae is that they offer us insights into the mass-loss history of the star, not just during the WR stage, but going back to any earlier episodic mass loss events. The fact that amateur images like those presented here on WR 71 and on other WR nebulae like WR 8 \citep{Fesen2025} suggests that substantially deeper imaging reconnaissance of WR star nebulae compared to that of earlier surveys may lead to additional WR ring nebula detections. Moreover, the inclusion of \O3 imaging is now realized to be important for detecting a WR star's complete associated optical nebulosity.

The current low detection percentage of WR stars with surrounding optical nebulae raises
the question of why only a fraction of WR stars show optical ring nebulae? Presumably this is partially a reflection of the star's local interstellar environment. Yet, on the other hand as this study has shown, prior imaging surveys may simply not have gone deep enough to detect very faint surrounding WR emissions, or were limited to H$\alpha$ emission thereby missing associated \O3 emission like that seen here for WR 71. A deeper and broader emission line imaging survey of WR stars might prove fruitful thereby enhancing our understanding on the frequency and formation of WR ring nebulae. \\

We thank Phil Massey and an anonymous referee for helpful suggestions that improved the paper. 

\facilities{Observatorio El Sauce, Chile}
\software{Photoshop, PixInsight, Astropixel 
Processor, DS9 fits viewer \citet{Joye2003}, WCSTools \citet{Laycock2010}  }
 

\bibliography{AAS_ref_5}{}
\bibliographystyle{aasjournal}

\end{document}